\documentclass[preprint]{revtex4}

\usepackage{amsfonts}
\usepackage{amsmath}
\usepackage{graphicx}

\usepackage{braket}

\newcommand{\w}{\omega}
\newcommand{\s}{\mathbf{s}}
\newcommand{\sn}{{\mathbf{s^0}}}
\newcommand{\Ds}{\mathbf{\Delta s}}
\newcommand{\f}{\mathbf{f}}
\newcommand{\E}{{\boldsymbol{\mathcal{E}}}}

\newcommand{\D}{\mathbf{D}}
\newcommand{\p}{\mathbf{p}}
\newcommand{\A}{\mathbf{A}}
\newcommand{\R}{\mathbf{r}}

\begin{document}

\title{Coulomb time delays in high harmonic generation}
\author{Lisa Torlina$^{1}$ and Olga Smirnova$^{1}$,$^{2}$}
\email{olga.smirnova@mbi-berlin.de}
\address{$^{1}$Max Born Institute for Nonlinear Optics and Short Pulse Spectroscopy, Max-Born-Strasse 2 A, 12489 Berlin, Germany,$^{2}$ Technische Universit{\"a}t Berlin,
Ernst-Ruska-Geb{\"a}ude, Hardenbergstr. 36A,10623, Berlin, Germany}

\begin{abstract}
Measuring the time it takes to remove an electron from an atom or molecule during photoionization using newly developed attosecond spectroscopies has been a focus of many recent experiments.
However, the outcome of such measurement depends on measurement protocols and specific observables available in each particular experiment. One of such protocols relies on high harmonic generation.
First, we derive rigorous and general expressions for ionization and recombination times in high harmonic generation experiments. We show that these times are different from, but related to ionization times measured in photo-electron spectroscopy, i.e. using attosecond streak camera, RABBITT and atto-clock methods. 
Second, we use the Analytical R-Matrix theory (ARM) to calculate these times and compare them with experimental values.
\end{abstract}
\maketitle

\section{Introduction}

The problem of time-resolving the removal of an electron during photoionization 
and studying time-delays associated with this process is an intriguing one \cite{pazourek2015attosecond}. Besides the fundamental implications for our understanding of atom-light interaction, such measurements have 
the capacity to serve as a sensitive probe of multielectron dynamics \cite{carette2013multiconfigurational,Torlina2015} and have an important role to play in calibrating attosecond recollision-based pump-probe  experiments \cite{Shafir2012}. 

Indeed, recent experimental developments 
-- including the attosecond streak camera \cite{Schultze2010}, the attoclock \cite{Eckle2008a,Eckle2008}, attosecond transient absorption \cite{Goulielmakis2010}, RABBIT \cite{klunder2011probing,guenot2012photoemission,Guenot2014} and high harmonic spectroscopy \cite{Shafir2012,soifer2013spatio,Pedatzur2015}, -- 
have made it possible to measure ionization times down to the level of tens of attoseconds, both in the one-photon 
\cite{Schultze2010,klunder2011probing,guenot2012photoemission} 
and multi-photon regimes
\cite{Eckle2008a,Eckle2008,Goulielmakis2010,pfeiffer2012attoclock,Shafir2012,soifer2013spatio,Pedatzur2015}. 
In each case, however, a thorough theoretical understanding \cite{Madsen2012,dahlstrom2013theory,Ivanov2011,Nagele2011,serbinenko2013multidimensional,Maquet2014,Gaillac2016} of the underlying physics has been crucial to correctly interpret the experimental results. 
To extract time information from the data, it is necessary to formulate a connection between the classical concept of time, the quantum wavefunction and the experimental observables. The reconstructed times are inherently very sensitive to the way in which this is done.

In the one-photon case, the theoretical basis for doing so
is now well-established (e.g. see \cite{Dahlstrom2012,pazourek2015attosecond}). 
Ionization is thought of as a half-scattering process and time delays are defined in terms of the Eisenbud-Wigner-Smith (EWS) time \cite{Wigner1955,Smith1960}: the derivative of the scattering phase-shift of the photoelectron with respect to its energy
\begin{equation}\label{eq:EWStime}
	\Delta t_{WS} = -\frac{d\phi}{dE}.
\end{equation}

In the multiphoton regime, however, the situation is notably less straightforward. If we think of ionization as a tunnelling process, we are presented with a number of different possible definitions for the tunnelling time, 
and it is not clear from the outset which is the `correct' time to use (e.g. see \cite{Landauer1994,DeCarvalho2002,Yamada2004}).
What's more, the use of any of these definitions in the context of strong field ionization necessarily restricts us to the tunnelling limit. 
It is not clear 'a-priori' how to take non-adiabatic effects into account or investigate how time delays are affected as we move towards the few-photon limit \cite{Kaushal2015}.

Alternatively,
the concept of ionization time arises in another way within certain analytical approaches to strong field ionization. 
These include the widely used and broadly successful strong field approximation (SFA) -- which we use here as an umbrella term for the  related work of Keldysh \cite{Keldysh1965}, Perelomov, Popov and Terent'ev \cite{PPT1,PPT2}, Faisal \cite{Faisal1973} and Reiss \cite{Reiss1980} -- as well as the more recently developed analytical R-matrix (ARM) method \cite{Torlina2012,Torlina2012b,Kaushal2013,Kaushal2015b, Torlina2013,Torlina2014ab}. The latter is a fully quantum theory that is able to accurately describe the long-range interaction between the outgoing electron and the core -- the absence of which is the main limitation of the SFA. 
In each of these approaches, it is necessary to integrate over a time variable that describes the instant at which the bound electron first interacts with the field. In analytical approaches, the integral is typically evaluated using the saddle point method, and the real part of the complex saddle point solution $t_i=\mathrm{Re}[t_\iota]$ can then be interpreted as the most probable time of ionization -- that is, the time at which the electron appeared in the continuum. The concept of saddle point time as ionization time also features in the Coulomb-corrected SFA approach (CCSFA) \cite{Popruzhenko2008a,Popruzhenko2008b} (see also recent review \cite{popruzhenko2014keldysh} and recent update of the theory \cite{popruzhenko2014invariant}), which takes the SFA ionization amplitude as its starting point 

Recently, the above idea was applied to analyse \cite{Torlina2015} the results of the attoclock experiments \cite{Eckle2008a,Eckle2008,pfeiffer2012attoclock}. The basic premise here relies on using short pulses of circularly or nearly-circularly polarized light to induce ionization and deflect electrons in different directions, depending on their time of ionization \cite{Eckle2008a,Eckle2008}. In order to reconstruct the ionization time $t_i$, it is necessary to establish its relationship to the observed angle and momentum at the detector $(\theta,p)$. This was possible in ARM using the saddle point analysis, yielding the following expression \cite{Torlina2015}:
\begin{equation}
	t_i = 
	\left( \frac{\theta}{\w} + \Delta t^\mathrm{env}(\theta,p) \right) + \Delta t^C(\theta,p),
\end{equation}
where 
\begin{equation}\label{eq:tauC}
	\Delta t^C = - \frac{\partial \phi^C}{\partial I_p}, 
\end{equation}
$\phi^C$ is the phase accumulated by the outgoing electron due to its interaction with the ionic core,
$I_p$ is the ionization potential of the bound state from which the electron escaped, and 
$\Delta t^\mathrm{env}(\theta,p)$ is a small correction due to the shape of the pulse envelope. Expression \ref{eq:tauC} has also been derived within the ARM method in Ref. \cite{Kaushal2015b}. Applying this relationship to the results of ab initio numerical experiments, it was possible to show that there are no time delays associated with the tunnelling process itself, at least for the simplest case of the hydrogen atom. However, delays $\Delta t^C$ due to the electron-core interaction could be important and generally cannot be neglected, especially in the case when two-electron excitations lie close to the first ionization threshold \cite{Torlina2015}. 

In fact Eq.\eqref{eq:tauC}, which we can think of as the delay accumulated in a long-range potential compared to a short-range potential, coincides with an expression for ionization delay derived in an entirely different way. In \cite{Kaushal2015}, the idea of the Larmor clock -- originally proposed in the context of tunnelling times -- was applied to ionization. It was shown to reduce to the EWS time in the single-photon limit and reproduce Eq.\eqref{eq:tauC} in the strong field regime. Thus, the different definitions for times are not unrelated: there is a link between the saddle point-based ionization time, the Larmor tunnelling time and the EWS single-photon delay.

The attoclock, however, is not the only means by which ionization times can be measured in the strong field regime. Two-colour high harmonic spectroscopy experiments offer another elegant and powerful approach to this problem, using a weak probe pulse to perturb the ionization dynamics in a controlled way \cite{Shafir2012}. In light of this, a number of questions naturally arise. Are the times measured in HHG the same as those measured by the attoclock? If not, how are they related? How does the electron-core interaction imprint itself on the ionization and recombination times in this case? 

Here, we address these questions by extending the saddle point analysis discussed above to HHG. After briefly reviewing the basic approaches for describing time in HHG in Section \ref{sec:timeinHHG}, we present a general analysis that incorporates electron-core interaction in Section \ref{sec:generalanalysis}. In Section \ref{sec:calcinARM}, we discuss how the calculation can be implemented in practice using ARM. We present and discuss the results of such a calculation in Sections \ref{sec:results},\ref{sec:results1}. Section \ref{sec:conc} concludes the work.


\section{Ionization and recombination times in HHG: a brief overview}\label{sec:timeinHHG}

\subsection{The classical model}

The simplest theoretical description of ionization time in the context of HHG comes from the classical model, which describes the process in terms of three steps: tunnel ionization, classical propagation in the continuum 
and recombination. In the standard classical, so-called three-step model \cite{Corkum1993}, (see also  pertinent refs. \cite{schafer1993above,multiphoton1988,corkum1989above,kuchiev1987atomic}) electron-core interaction is neglected during the classical propagation step, and it is assumed that the electron starts its continuum motion with a velocity of zero. If we also assume that recombination occurs when the electron returns to its starting point and relate its kinetic energy at this instant with the energy of the emitted photon, we obtain the following three equations:
\begin{align}
	& \mathbf{v}(t_i) = 0 \label{sm1}\\
	& \int_{t_i}^{t_r} \mathbf{v}(t) dt = 0 \label{sm2}\\
	& \frac{1}{2}v(t_r)^2 = E_\w - I_p \label{sm3},
\end{align}
where $\mathbf{v}(t) = \mathbf{p} + \mathbf{A}(t)$ is the electron velocity, $\mathbf{A}$ is the vector potential that describes the laser field $\mathcal{E}=-\frac{\partial \mathbf{A}}{\partial t}$, and $E_\w = N \w$ is the energy of the emitted photon.

For any given photon energy $E_\w$, the above equations can be solved for the ionization time $t_i$, the recombination time $t_r$ and the canonical momentum $\mathbf{p}$. As such, we can associate an ionization and recombination time to each harmonic number: this defines the mapping between experimental observable and time in this case. However, although excellent as a first approximation, this model is clearly rather crude. It artificially matches a quantum mechanical description of the ionization step with  classical propagation, assumes very simple initial conditions for the electron's continuum motion and ignores the influence of the positively charged ionic core. 
It is not surprising, therefore, that a comparison of the above predictions with times obtained in high harmonic spectroscopy experiments revealed a notable discrepancy \cite{Shafir2012}.

\subsection{The strong field approximation (SFA)}

As mentioned in the introduction, the SFA offers a more sophisticated quantum approach to the problem of time in strong field ionization, which is based on saddle point analysis. In essence, the key approximation of the SFA is to neglect the interaction between the electron and its parent ion after the instant ionization or, equivalently, to assume that the core potential is short range. 
Doing so, the induced dipole can be expressed as \cite{lewenstein1994theory}: 
\begin{equation}\label{SFAintegral}
	\D(N\w) = -i \int dt \int d\p \int dt'
	\ P(t',t,\p)	
	\ e^{i E_\w t} \ e^{- i I_p (t-t')} 
	\ e^{-i S_V(t,t',\p)},
\end{equation}
where $S_V$ is the Volkov phase
\begin{equation} \label{eq:Sv}
	S_V = \frac{1}{2} \int_t^{t'} v^2(\tau) \ d \tau,
\end{equation}
and $P$ is a prefactor that varies relatively slowly.
The times $t'$ and $t$ over which we integrate can be associated with ionization and recombination respectively.

The presence of a large phase, which leads to rapid oscillations of the integrand, makes it possible to evaluate the above integral using the saddle point method. It tells us that the integral will be accumulated predominantly in the vicinity of points where the derivative of the phase vanishes, which are defined by the saddle point equations \cite{lewenstein1994theory,salieres2001feynman}:
\begin{align}
	& \frac{\partial S_V}{\partial t'}
	= I_p \label{SFA1}\\
	& \frac{\partial S_V}{\partial \mathbf{p}} 
	= 0 \label{SFA2}\\
	& \frac{\partial S_V}{\partial t} 
	= E_\omega - I_p \label{SFA3}.
\end{align}
We shall denote the solutions to these equations by $t_\iota^0,\p_s^0,t_\rho^0$.

In fact, it is easy to check that Eq.\eqref{SFA2} and Eq.\eqref{SFA3} coincide exactly with Eq.\eqref{sm2} and Eq.\eqref{sm3} from the classical model, while Eq.\eqref{SFA1} is a modified version of Eq.\eqref{sm1}: $v^2(t') = - 2 I_p$. 
The latter, in fact, gives rise to a key difference between the two descriptions. Whereas the solutions in the classical model are fully real, their counterparts in SFA are complex in general (see e.g. discussion in \cite{smirnova2014multielectron,smirnova2013multielectron}, ). Nevertheless, as mentioned in the introduction, the real parts of the saddle point solutions come with an interpretation. (This interpretation is particularly transparent in case of short-range potentials, where  similar analysis yields excellent agreement with numerical calculations for attoclock \cite{Torlina2015}).  We can associate $t_i^0 = \mathrm{Re}[t_\iota^0]$, $\p^0 = \mathrm{Re}[\p_s^0]$ and $t_r^0 = \mathrm{Re}[t_\rho^0]$ with the time of ionization, canonical momentum and time of recombination respectively. As such, Eq.\eqref{SFA1}-\eqref{SFA3} again define a mapping between time and harmonic number, albeit a somewhat different one.

Using this model, it was shown that agreement with times reconstructed from high harmonic spectroscopy experiments is notably improved \cite{Shafir2012}. However, what both the classical model and the SFA have in common is the neglect of the electron-core interaction throughout the electron's motion in the continuum. 
In the context of the attoclock, we have seen that this is not sufficient: ignoring the influence of the core potential on saddle point solutions leads to qualitatively and quantitatively incorrect results \cite{Torlina2015}. Motivated by this, let us now consider how the above solutions are modified if we allow for an electron-core interaction term.

\section{Influence of electron-core interaction on times in HHG: a general analysis}\label{sec:generalanalysis}

Within ARM, we know that the SFA expression for the ionization amplitude -- the analogous quantity to the induced dipole given by Eq.\eqref{SFAintegral} -- is modified by the addition of a Coulomb phase $e^{-i W_C}$ \cite{Torlina2012}. Let us now consider what a term of this nature would imply for times in HHG. 

In particular, suppose that the integral in Eq.\eqref{SFAintegral} now includes some additional factor $e^{-iF(t',t,\p)}$, which accounts for the interaction between the active electron and the core. Although ARM provides us with an explicit expression for $F$, which we shall return to in Section \ref{sec:calcinARM}, we shall keep the analysis general for the time being. We assume only that $F$ is small compared to $S_V$ and allow it to be complex in general. The real part of $F$ then specifies the phase, which we shall denote by $\phi$.

The saddle point equations \eqref{SFA1}-\eqref{SFA3} are modified as follows 
\begin{align}
	& \frac{\partial S_V}{\partial t'} 
	+ \frac{\partial F}{\partial t'} = I_p \label{CC1}\\
	& \frac{\partial S_V}{\partial \p} 
	+ \frac{\partial F}{\partial \p} = 0 \label{CC2}\\
	& \frac{\partial S_V}{\partial t} 
	+ \frac{\partial F}{\partial t} = E_\omega - I_p \label{CC3},
\end{align}
and since $F$ is small compared to $S_V$, we can search for solutions of the form $(t_\iota^0 +\Delta t_\iota, p_s^0 +\Delta p_s, t_\rho^0 +\Delta t_\rho)$, where $(t_\iota^0, p_s^0, t_\rho^0)$ satisfy Eq. \eqref{SFA1}-\eqref{SFA3}. 
Expanding about the SFA saddle points, keeping only first order terms in $F$, and using the chain rule to rewrite the derivatives, we arrive at the following result (see Appendix): 
\begin{align}
		\Delta t_\iota &= -\frac{\partial F}{\partial I_p} 
			-\frac{\partial F}{\partial E_\w} \label{eq:dts}\\
		\Delta t_\rho &= -\frac{\partial F}{\partial E_\w}. \label{eq:dtr}
\end{align}
In general, since $F$ is complex, these corrections will also be complex. However, as before, we can assign an interpretation to their real parts:
$\Delta t_i = \mathrm{Re}[\Delta t_\iota]$ and $\Delta t_r = \mathrm{Re}[\Delta t_\rho]$ encode the delays due to the electron-core interaction imprinted upon ionization and recombination times respectively:
\begin{align}
		\Delta t_i &= -\frac{\partial \phi}{\partial I_p} 
			-\frac{\partial \phi}{\partial E_\w} \label{eq:dtsr}\\
		\Delta t_r &= -\frac{\partial \phi}{\partial E_\w}. \label{eq:dtr_real}
\end{align}

There are a few things worth noting about the above results. First, notice that the expression for the correction to the recombination time, $\Delta t_r = -\partial \phi/\partial E_\omega$, is reminiscent of the EWS time given by Eq.\eqref{eq:EWStime}. Indeed, we can understand this if we recall that recombination is simply single-photon ionization run in reverse. Varying with respect to the photon energy is directly equivalent to varying with respect to the kinetic energy of the recombining electron. Second, note that the correction to ionization time in HHG, $\Delta t_i = - \partial \phi/\partial I_p - \partial \phi/\partial E_\w$, has an extra term compared to its counterpart in strong field ionization given by Eq.\eqref{eq:tauC}. Subtracting the above expressions for $\Delta t_r$ and $\Delta t_i$ yields
\begin{equation}\label{eq:dphidIpHHG}
	\Delta t_r - \Delta t_i = \Delta(t_r-t_i) = \frac{\partial \phi}{\partial I_p}.
\end{equation}	
In other words, the derivative of the phase with respect to the ionization potential in HHG tells us how much more (or less) time the electron will spend in the continuum before it recombines, as a consequence of electron-core interaction.

\section{Coulomb corrections in HHG using ARM}\label{sec:calcinARM}

Keeping the function $F$ unspecified, this is as far as we can go -- if we want to determine the corresponding time delays in practice, we must evaluate $F$ explicitly. Luckily, this is precisely what ARM allows us to do.

In particular, ARM tells us that in strong field ionization $e^{-iF}$ is replaced by $e^{-iW_C^\mathrm{SFI}}$, where
\begin{equation}\label{eq:WCSFI}
	W_C^\mathrm{SFI}(\p,T) = \int_{t_\kappa}^{T} dt' \ U(\R_s(\p,t')),
\end{equation}
$\p$ is the electron momentum measured at the detector, $T$ is the time of observation, $U(\R)$ is the core potential and $\R_s$ is the Coulomb-free electron trajectory, 
\begin{equation}
	\R_s(\p,t') = \int_{t_\iota}^{t'} (\p + \A(t'')) dt''.
\end{equation}
$W_C^\mathrm{SFI}$ is the electron action assosiated with the Coulomb-laser coupling \cite{Smirnova2008,smirnova2007coulomb}. The lower limit of the integral in Eq.\eqref{eq:WCSFI}, $t_\kappa = t_\iota - i/\kappa^2$, is determined by the boundary-matching procedure for the outgoing electron \cite{Torlina2012,Kaushal2015b}. 

The purpose of the matching procedure is to avoid using $W_C^\mathrm{SFI}$ outside the limits of its applicability range, i.e. close to the core. This procedure enables a smooth merger of the asymptotic tail of bound electron wave-function with the quasicalssical wave-function of the escaping electron, driven by the laser field. The dominant fraction of electrons liberated by strong field ionization arrive to the detector without revisiting the core and therefore the matching is done only once, when electron departs from the core. However, in our case we have to consider these, not-very-likely return events, since only they give raise to HHG.
It means that we need to perform the matching procedure once again, when electron returns back to the core, linking the phase accumulated due to Coulomb-laser coupling between ionization and recombination to the field-free continuum solution for the returning electron.

Fortunately, an equivalent boundary-matching problem has already been solved. In \cite{Ivanov2011}, single photon ionization in the presence of a probing infrared field was analysed in the context of the attosecond streak camera. There, a matching argument was used to show that the effective starting point for an electron trajectory with initial velocity $v_0$ is given by 
\begin{equation}
	r_0(v_0) = \frac{1}{v_0 \ a(v_0)}, \label{eq:r0}
\end{equation}
where 
\begin{equation}\label{eq:a}
	a(v_0) = 2 e^{-2 \gamma_E} e^{2 \xi(v_0)},
\end{equation}
$\gamma_E$ is Euler's constant and 
\begin{equation}\label{eq:xi}
	\xi(v_0) = \sum_{n=1}^\infty \frac{1}{n} \left[1 - v_0 n \arctan\left(\frac{1}{v_0 n}\right)\right].
\end{equation}

Noting that recombination in HHG is simply the reverse of this process (that is, the emission (rather than absorption) of a photon in the presence of an infrared field), we can apply this result directly to determine $t_\mathrm{end}$. In particular, we now think of $r_0$ as the end point of our electron trajectory and set $r_0 = r_s(t_\mathrm{end})$. The corresponding velocity $v_0$ is the velocity at recombination, 
\begin{equation}\label{eq:vr}
	v_0 = v_r = \sqrt{2 (E_\w - I_p)}.
\end{equation}
For any given photon energy $E_\w$, we then have
\begin{equation}\label{eq:rseqr0}
	r_0(v_r) = r_s(t_\mathrm{end}) = \int_{t_\iota(E_\w)}^{t_\mathrm{end}} (p_s(E_\w) + A(t))\ dt,
\end{equation}
which can be used to solve for $t_\mathrm{end}$, using Eq.\eqref{eq:r0}-\eqref{eq:vr} to evaluate $r_0(v_r)$.

\begin{figure}[h!]
  \begin{center}
    \includegraphics[width=0.5\textwidth]{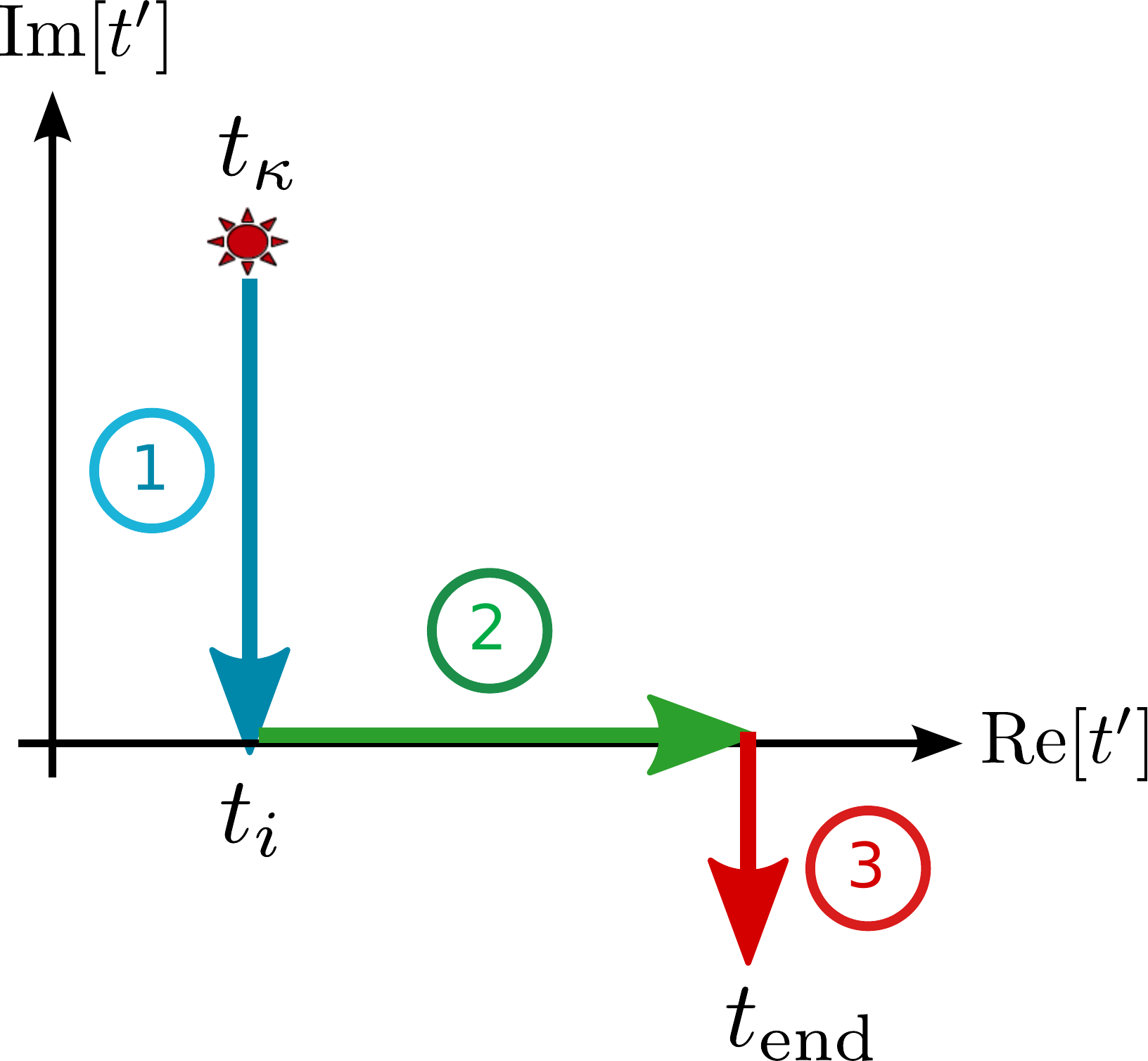}
      \end{center}
  \caption{Contour for the $W_C$ integral in HHG.}
  \label{fig:contour}
\end{figure}

In doing so, it should be noted that $r_s(t)$ is complex in general for real times, whereas $r_0$ is always real. Consequently, in order to satisfy the above equation, we must allow $t_\mathrm{end}$ to be complex as well. This tells us that, in contrast to ionization, our integral for $W_C$ no longer ends on the real axis: both start and end points are now complex. The generalization, however, is straightforward. When describing Coulomb effects in strong field ionization, the integration contour was chosen in two parts: first, down from $t_\kappa$ to $\mathrm{Re}[t_\iota]=t_i$ on the real axis, and then along the real axis up to time $T$ \cite{Torlina2012,Torlina2013}. These two legs were interpreted in terms of tunnel ionization and the electron's motion in the continuum respectively, following PPT \cite{PPT2}. For HHG, we simply add a third leg: down from $\mathrm{Re}[t_\mathrm{end}]$ on the real axis to $t_\mathrm{end}$ (see Fig.\ref{fig:contour}).

We note that, in general, the interpretation of the real part of the complex saddle point  as the ionization time is not connected to a certain choice of the integration contour.  Analytic properties of the integrands in $W_C$, of course, allow  a considerable deformation of the contour without any influence on the result of the integration. 
In contrast to the contour, 
the saddle point in time itself is unique and well defined; it has real and imaginary part. The real part of the saddle point can also be directly detected in the numerical  attoclock experiments \cite{Torlina2015}. In the long wave-length limit,  the attoclock observable –- the so-called off-set angle -- is equal to the real part of the saddle point \cite{Kaushal2015}.
What's more, in Ref \cite{Kaushal2015} we have introduced an alternative method for deriving the ionization time. This method uses neither the concept of tunneling, nor the  concept of trajectories, nor does it rely on the saddle method. The result for the ionization time remains  the same.


To extend Eq. \ref{eq:WCSFI}  to describe HHG, there are only two major changes we need to make: 1. the momentum at the detector $\p$ should be replaced by the saddle point solution $\p_s(E_\w)$, and 2. the observation time $T$ should be replaced by the time $t_\mathrm{end}$. That is,
\begin{equation}\label{eq:WCHHG}
	W_C^\mathrm{HHG}(E_\w) = \int_{t_\kappa}^{t_\mathrm{end}} dt' \ U(\R_s(E_\w,t')),	
\end{equation}
where 
\begin{equation}
	\R_s(E_\w,t') = \int_{t_\iota(E_\w)}^{t'} (\p_s(E_\w) + \A(t'')) dt''.
\end{equation}
 We can readily determine the value of $\p_s(E_\w)$ from Eq.\eqref{SFA1}-\eqref{SFA3}.


\section{Results: Coulomb time delays in HHG}\label{sec:results}

Having determined $t_\mathrm{end}$ and chosen a contour, we have all the ingredients we need in order to evaluate the correction $W_C$ as given by Eq.\eqref{eq:WCHHG}. In itself, this tells us the first order effects on HHG spectra due to the electron-core interaction. However, as we saw in Section \ref{sec:generalanalysis}, we need one further step to learn about times: we must differentiate the Coulomb phase $W_C$ with respect to $I_p$ and $E_\w$ to find the corrections to the saddle point solutions. In practice, this can be done numerically by evaluating $W_C$ for two or more closely spaced values of $I_p$, $E_\w$.

We shall now compare these results with the times reconstructed from high harmonic spectroscopy measurements \cite{Shafir2012}. The reconstruction procedure is described in Ref. \cite{Shafir2012} (see also SI of  \cite{Shafir2012}) and analysied in detail in Ref. \cite{serbinenko2013multidimensional}. For the benefit of the reader here we briefly outline the main idea of the reconstruction.

Suppose HHG is driven by a strong laser field at the fundamental frequency $\omega$ and is described by the vector potential: $A_{\omega}(t)=\vec{e_x} A_0\sin(\omega t)$. The idea of detection of the ionization time is very simple.   It relies on the application of an additional perturbative control field at the frequency $2\omega$, $A_{2\omega}(t)= \vec{e_y} A_0\sin(2\omega t+\phi)$ phase locked to the fundamental field, and polarized in orthogonal direction to the fundamental field. This control field modulates HHG yield as a function of the relative phase $\phi$  between the two fields, i.e. as a function of the two-color delay $\phi$. This field “kicks” the electron in lateral direction, once it leaves the bound atomic state and exits from the tunneling barrier at the ionization time $t_i$. The magnitude and the sign of this kick is controlled by the two-color delay. The HHG signal maximizes for a specific two-color delay, $\phi_{max}$, when the lateral kick is equal to zero, i.e. when the electron displacement between ionization and recombination is minimized. To a good approximation, this requires the vector potential of the control field at the moment of ionization to be close to zero, $A_{2\omega}(t_i)= \vec{e_y} A_0\sin(2\omega t_i+\phi_{max})\approx 0$, thus the ionization time is $|t_i|\approx \phi_{max}/\omega$. Corrections accounting for the kick during tunneling are accounted for in the full reconstruction. Note that the second harmonic field also breaks the symmetry in electron dynamics  between the two consecutive 
laser half cycles, leading to the generation of even harmonics. 
The asymmetry maximises for those values of  $\phi$, for which 
the lateral velocity of the electron upon recombination is maximal. 
Recombination times are reconstructed following the maximal HHG signal for 
even harmonics, as suggested and performed in Ref. \cite{Shafir2012} and augumented in Ref. \cite{Lein2013} by including complex, rather than real SFA recollision times.  

Calibrating  $\phi_{max}$ in experiment allows one to reconstruct absolute values of ionization and recombination times in HHG. If the phase is not calibrated, the experiment can only reconstruct (i) the delay between ionization and recombination time, (ii) the dependence of the 
ionization and recombination times on harmonic energy. To compare our results 
with the experiment we shift both ionization and recombination times by $-20$ asec, uniformly for 
all harmonics.

\begin{figure}[h!]
  \begin{center}
    \includegraphics[width=\textwidth]{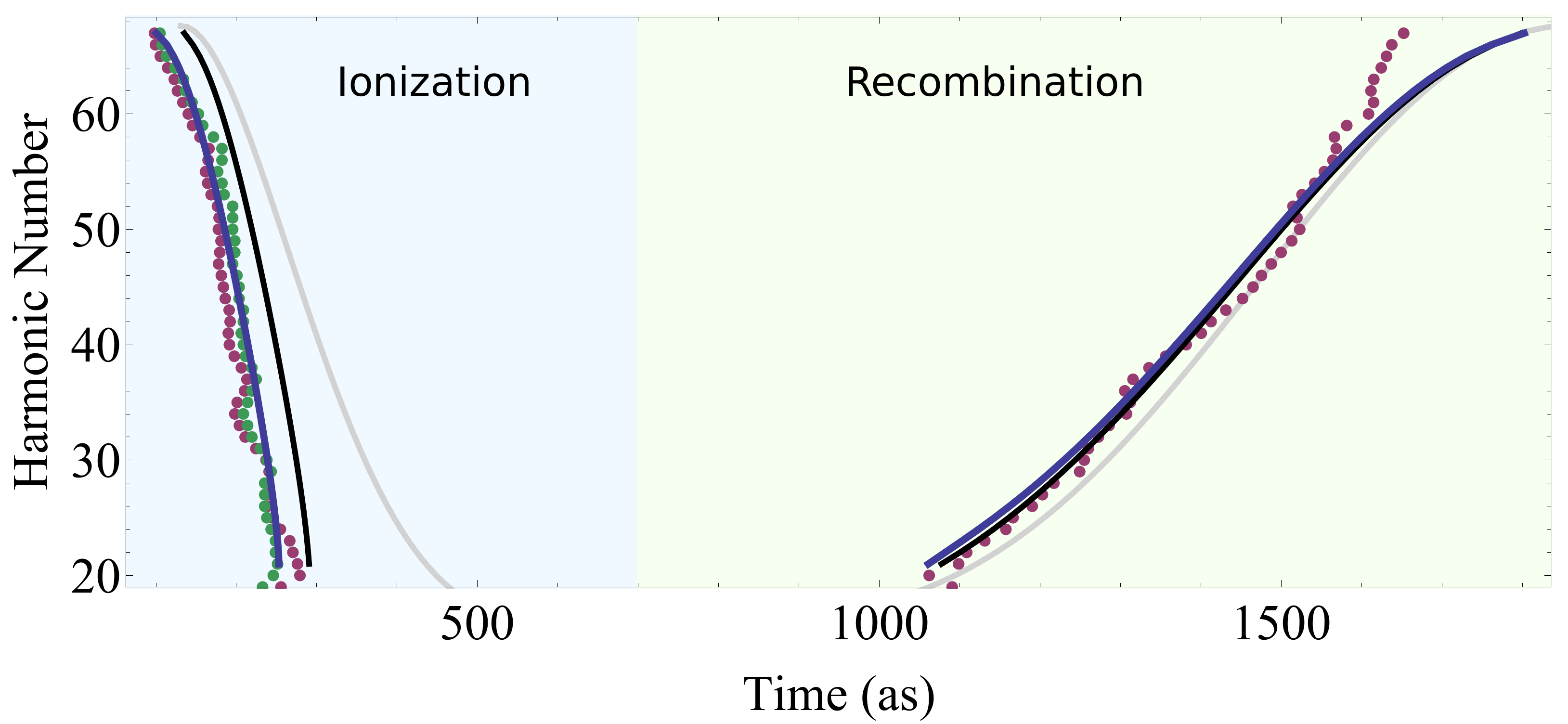}
      \end{center}
  \caption{Ionization and recombination times predicted using the classical model (grey lines), SFA (black lines) and ARM (blue lines) compared with times reconstructed from high harmonic spectroscopy experiments (pink and green dots) (see \cite{Shafir2012}) for helium atoms at $\lambda=800$nm and $I \approx 3.8 \times 10^{14}$ W/cm$^2$.}
  \label{fig:theoryexpt}
\end{figure}

Figure \ref{fig:theoryexpt} shows the results of applying this procedure for the helium atom. The  ionization and recombination times obtained using ARM (blue lines) are compared to SFA (black lines), the classical model (grey lines) and times reconstructed from high harmonic spectroscopy experiments (pink and green dots, originally published in \cite{Shafir2012}). Compared to the SFA, we find that the Coulomb-corrected ionization times are shifted to earlier values. This effect has only a weak dependence on on harmonic number: the shift varies between $\sim 33$ and $\sim 37$ attoseconds, decreasing slightly with $N$. The recombination times are notably less affected overall, though they display a stronger dependence on the harmonic number: the shift in this case is between $\sim 5$ and $\sim 19$ attoseconds, again decreasing with $N$. Putting these two facts together, we see that the total amount of time the electron spends in the continuum (given by $\partial \phi/\partial I_p$) increases by $\sim$ 18-28 attoseconds. 

Comparing our results with times reconstructed from high harmonic spectroscopy measurements, we find that ARM offers a notable improvement over the SFA, where electron-core interaction was neglected. Although the corrections to ionization and recombination times are only of the order of tens of attoseconds, they are nevertheless clearly within the resolution of current state-of-the-art HHG experiments.

We note that our method also allows us to analyse the Coulomb corrections to the 
imaginary times. This has been done in Ref.\cite{Pedatzur2015}, where 
 imaginary ionization times where reconstructed 
from the experimental measurements.

\section{Why Coulomb time delays are so small?}
\label{sec:results1}
We have derived the times in HHG using the saddle point method and iterative procedure for finding saddle point solutions, thus the derived times correspond to solutions of saddle point equations. Iterations treat the electron action associated with the motion in the Coulomb field (so called Coulomb-laser coupling term) as a perturbation to the action due to the laser field.   The zero-order iteration yields the well-known SFA (Strong Field Approximation) results for the times, when the Coulomb-laser coupling is neglected.  
To obtain a small parameter of the iterative procedure, we will compare the results of subsequent iterations: $|\Delta t_\iota|$/ $|\Delta t_\iota^0|$, where $|\Delta t_\iota|$ is the complex ionization time in the first order and $|\Delta t_\iota^0|$ is the complex  ionization time in the SFA (zero-order iteration).

The Coulomb correction to ionization time in HHG has two components $\Delta t_\iota =t_\iota^{SF}+t_{WS}^{SF}$:
\begin{align}
		t_\iota^{SF} &= -\frac{\partial F}{\partial I_p} ,\\
		t_{WS}^{SF}&=	-\frac{\partial F}{\partial E_\w} \label{eq:dts1}.
\end{align}
First, we discuss the small parameter for the first term.
To obtain simple analytical results for the small parameter we consider the low-frequency (tunnelling) limit. In this case $|\Delta t_\iota^0|=\tau$, where  $\tau=\frac{\sqrt{2I_p}}{\mathcal E_0}$ is the well-known tunnelling time (the imaginary component of the saddle point solution), $\mathcal E_0$ is the strength of the laser field. The real part of the saddle point  can be chosen to be zero in the tunnelling limit. Now we need to discuss the absolute value of complex first order correction to saddle point time. 
One can obtain a very simple analytical expression for the real part of this correction: $t_\iota^{SF}\approx Z/I_p^{3/2}$, where $Z$ is the charge of the core, $I_p$ is ionization potential  (see derivation in  Refs \cite{Kaushal2013,Kaushal2015b}). Note that the absolute value of the saddle point solution is determined by its real component, because the correction to the imaginary component is zero in the first order iteration, in the tunnelling limit.
We obtain $\zeta_1=\frac{|t_\iota^{SF}|}{|t_\iota^{0}|}
=n^{*}\frac{3\mathcal E_0}{2 \mathcal E_c} \frac{2^{5/2}}{3}=
\frac{n^*}{2ImS_{SFA}}\frac{2^{5/2}}{3}$,  where $n^*=Z/ I_p^{1/2}$is the effective principal quantum number of a given quantum state and $ImS_{SFA}$  is the imaginary part of the electron action associated with its dynamics in the laser field only. Note that $2ImS_{SFA}=\frac{2\mathcal E_c}{3 \mathcal E_0}$  is the famous exponent that appears in the equation for the tunnelling rate: $\Gamma=exp(-2ImS_{SFA})$. It is well known that this formula is only applicable if $ImS_{SFA} >>1$. 
Thus, we have our small parameter $\zeta_1=\frac{n*}{ImS_{SFA}}\frac{2^{3/2}}{3}$. The quantum number $n^*$ characterizes the action of the electron in the bound state, while $S_{SFA}$ characterizes the action of the electron, driven by the laser field. If the action due to the laser-driven dynamics exceeds the action in the ground state (which is usually the case in strong field ionization), the ionization time $|t_\iota^{SF}|$ will remain small. This condition is essentially the same as the condition of applicability of most of the Coulomb-corrected SFA theories.

The strong field ionization time $|t_\iota^{SF}|$ is only  a first part of the expression for ionization time in HHG:  $\Delta t_\iota =t_\iota^{SF}+t_{WS}^{SF}$.  We now discuss the second contribution, which is also equivalent to the first order correction to recombination time. $t_{WS}^{SF}=	-\frac{\partial F}{\partial E_\w}$ is the analogue of the well-known  Wigner-Smith ionization time, albeit the phase  $F$ entering this expression is not the field-free scattering phase, but is the phase that includes the effects of the laser field and describes the Coulomb-laser coupling. As discussed in the previous section, $|t_{WS}^{SF}|$ is smaller than $|t_\iota^{SF}|$.
In the case $|t_{WS}^{SF}|$ we face an interesting effect of partial cancellation of the purely Coulombic (truly Wigner-Smith field free delays) with the delays induced by the laser field. This effect is described in detail in Ref. \cite{Ivanov2011}. The small parameter in this case is the shift of electron momentum $\Delta q$ due to Coulomb-laser coupling (see  \cite{smirnova2007coulomb}) with respect to velocity of the returning electron: $\zeta_2=\frac{\Delta q}{v_r}$, where $v_r=2^{1/2}(E_{\omega}-I_p)^{1/2}$, $E_{\omega}$ is the harmonic energy. Using the expression for the momentum shift $\Delta q$ due to the Coulomb-laser coupling  from \cite{smirnova2007coulomb}, we find $\zeta_2=\frac{Z|\mathcal E(tr)|}{ |v_r| |2^{3/2}(E_{\omega}-I_p)^ {3/2}|)}$.  Note that the expressions for small parameters $\zeta_1$ and $\zeta_2$ have similar structure: $\zeta_{1,2}=Z \mathcal E/v^4$, where for $\zeta_1$ we should use the electron velocity in the bound state $|v|= (2I_p)^{1/2}$ and the laser field at the moment of ionization, but for $\zeta_2$ we need to use the electron velocity  $v_r=2^{1/2}(E_{\omega}-I_p)^{1/2}$ and the laser field at the moment of recombination.
Thus, in general, Coulomb corrections to recombination times are smaller than Coulomb correction to ionization times due to larger velocity of the returning electron than departing electron and smaller value of instantaneous laser field at ionization than at recombination for majority of electron trajectories, except the ones corresponding to low harmonic numbers.

We note that small values of the Coulomb shifts in HHG times make their detection challenging. In particular, they have not been seen in numerical simulations of Ref. \cite{Lein2013}.

\section{Conclusions and outlook}
\label{sec:conc}
We have shown how ionization and recombination times in HHG are modified when the long-range interaction between the active electron and the ionic core is taken into account. The resulting corrections are closely related to the delays in strong field and single photon ionization respectively, though they are not identical. In particular, the expression for ionization delay in HHG, $\Delta t_i ^{HHG} = - \partial \phi/\partial I_p - \partial \phi/\partial E_\w$, contains an additional term (with respect to strong field ionization time $\Delta t_i^{SFI} = - \partial \phi/\partial I_p$, derived earlier in \cite{Torlina2015,Kaushal2015,Kaushal2015b}) that factors in the recombination step.
The origin of the "additional" ionization delay  is related to different measurement protocols used in HHG and strong-field ionization experiments. While the latter detects photoelectrons, the former detects photons, where both ionization and electron recombination precede the measurement.
Therefore, in case of  HHG the "delay-line" on the way to the detector is associated not only with ionization (which includes propagation in the continuum), but also with recombination, leading to production of XUV light that is further detected to extract ionization delays.

Comparing the predictions of the ARM theory -- in which the Coulomb interaction is accounted for -- with times measured in high harmonic spectroscopy experiments, we find that the agreement is excellent. The fit is visibly better than for the SFA, where such effects are omitted. Thus, although relatively small, we can conclude that the electron-core interaction leaves a measurable and distinct signature on times in HHG. As such, it should be taken into consideration when calibrating attosecond recollision-based pump-probe experiments and interpreting experimental data.

\section{Acknowledgements}
The authors  gratefully acknowledge the support of Deutsche Forschungsgemeinschaft, project SM 292/2-3. We thank H. Sofier and N. Dudovich for providing the experimental data from \cite{Shafir2012}, shown in Figure 2. We thank  V. Serbinenko, F. Morales and M. Ivanov for discussions.


%
%
%


\clearpage

%
%

\newpage

\appendix

\section{Appendix}

In this section we present the derivation of Eq.\eqref{eq:dts}-\eqref{eq:dtr}. To do so, it will be convenient to introduce the following vectors
\begin{align}
	& \s = (t',p,t) \\
	& \sn = (t_\iota^0, p_s^0, t_\rho^0) \\
	& \Ds = (\Delta t_\iota, \Delta p_s, \Delta t_\rho) \\
	& \E = (I_p,X,E_\w)
\end{align}
and the vector-valued function
\begin{align}
	\f: \quad & \mathbb{R}^3 \rightarrow \mathbb{R}^3 \nonumber \\ 
			& (I_p,X,E_\w) \mapsto (I_p,X,E_\w - I_p).
\end{align}
The gradient $\nabla$ will be defined as a row vector, and $\nabla_\sn F$ will occassionally be used as shorthand for $\nabla_\s F(\sn)$.

Using this notation, the saddle point equations \eqref{CC1}-\eqref{CC3} can be expressed succinctly as 
\begin{equation}\label{eq:fullspvec}
	\nabla_\s S_V(\sn + \Ds) + \nabla_\s F(\sn + \Ds) = \f(\E),
\end{equation}
while their analogues in the SFA become
\begin{equation}\label{eq:SFAspvec}
	\nabla_\s S_V(\sn) = \f(\E),
\end{equation}
where we have added a constant $X \rightarrow 0$ to the LHS of the second saddle point equation for convenience in both cases. 

If we now expand the solution to Eq.\eqref{eq:fullspvec} about $\sn$, we obtain (to first order in $\Ds$ and $\nabla_\s F$)
\begin{equation}
	\nabla_\s S_V(\sn) + \Ds \ [J_\s[\nabla_\s S_V](\sn)]^T + \nabla_\s F(\sn) = \f(\E),
\end{equation}
where $J_\s$ is the Jacobian with derivatives taken with respect to $\s$.
Using Eq.\eqref{eq:SFAspvec} and the fact that the Jacobian of a gradient is the Hessian, we have
\begin{equation}
	\Ds \ [H_\s[S_V](\sn)]^T = - \nabla_\s F(\sn).
\end{equation}
Since the Hessian is symmetric and invertible (in this case), we can rewrite this as
\begin{equation}\label{eq:Ds1}
	\Ds = - \nabla_\sn F \ [H_\sn[S_V]]^{-1}.
\end{equation}

In principle, if we could evaluate $\nabla_\sn F$, we would be done.
However, we do not have direct control over the value of the complex saddle point $\s$ when we do the calculation numerically, which makes this a difficult quantity to work with. Instead, what we can do is vary the parameters $\E$ and see how $F$ changes as a result -- this makes it possible to evaluate $\nabla_{\E} F$ numerically. With this in mind, we would like to rewrite Eq.\eqref{eq:Ds1} in terms of $\nabla_\E F$ instead of $\nabla_{\s} F$.

To do so, we note that Eq.\eqref{eq:SFAspvec} establishes a functional relationship between $\sn$ and $\E$. In principle, we could solve this equation to find $\sn (\E)$. Taking the gradient of $F$ with respect $\E$ and applying the chain rule gives
\begin{equation}
	\nabla_\E F(\sn(\E)) = \nabla_\s F(\sn) \ J_\E [\sn(\E)],
\end{equation}
so
\begin{equation}\label{eq:gradsW}
	\nabla_\sn F = \nabla_\E F \ [J_\E [\sn]]^{-1}.
\end{equation}

We now need only evaluate the Jacobian $J_\E [\sn]$. To do so, let us differentiate both sides of the SFA saddle point equation Eq.\eqref{eq:SFAspvec} with respect to $\E$:
\begin{equation}
	J_\E [\nabla_\s S_V(\sn(\E))] = J_\E [\f(\E)].
\end{equation}
Applying the chain rule again, we can rewrite this as
\begin{equation}
	J_\s [\nabla_\s S_V(\sn)] \ J_\E[\sn(\E)] = J_\E [\f(\E)],
\end{equation}
and so (again making use of the fact that the Jacobian of a gradient is the Hessian),
\begin{equation}
	J_\E[\sn] = H_\sn[S_V]^{-1} \ J_\E [\f].
\end{equation}

Taking the inverse and substituting this into Eq.\eqref{eq:gradsW} gives
\begin{equation}
	\nabla_\sn F = \nabla_\E F \ [J_\E [\f]]^{-1} \ H_\sn[S_V].
\end{equation}
Finally, this allows us to rewrite Eq.\eqref{eq:Ds1} as
\begin{equation}
	\Ds = - \nabla_\E F \ [J_\E [\f]]^{-1} \ H_\sn[S_V] \ [H_\sn[S_V]]^{-1},
\end{equation}
which simplifies to
\begin{equation}
	\Ds = - \nabla_\E F \ [J_\E \f]^{-1}.
\end{equation}

In our case, $J_\E \f$ is very simple:
\begin{equation}
	J_\E \f = 
	\begin{pmatrix}
		1 & 0 & 0 \\
		0 & 1 & 0 \\
		-1 & 0 & 1 \\		
	\end{pmatrix},
\end{equation}
and
\begin{equation}
	[J_\E \f]^{-1} = 
	\begin{pmatrix}
		1 & 0 & 0 \\
		0 & 1 & 0 \\
		1 & 0 & 1 \\		
	\end{pmatrix}.
\end{equation}
This allows us to write down a solution for $\Ds$ in terms of $\nabla_\E F$:
\begin{align}
		\Delta t_s &= -\frac{\partial F}{\partial I_p} 
			-\frac{\partial F}{\partial E_\w} \\
		\Delta p_s &= -\frac{\partial F}{\partial X} \\
		\Delta t_r &= -\frac{\partial F}{\partial E_\w} .
\end{align}

\bibliography{CoulombinHHG} 

\begin{thebibliography}{10}

\bibitem{pazourek2015attosecond}
R.~Pazourek, S.~Nagele, and J.~Burgd{\"o}rfer, ``Attosecond chronoscopy of
  photoemission,'' {\em Reviews of Modern Physics}, vol.~87, no.~3, p.~765,
  2015.

\bibitem{carette2013multiconfigurational}
T.~Carette, J.~Dahlstr{\"o}m, L.~Argenti, and E.~Lindroth,
  ``Multiconfigurational hartree-fock close-coupling ansatz: Application to the
  argon photoionization cross section and delays,'' {\em Physical Review A},
  vol.~87, no.~2, p.~023420, 2013.

\bibitem{Torlina2015}
L.~Torlina, F.~Morales, J.~Kaushal, I.~Ivanov, A.~Kheifets, A.~Zielinski,
  A.~Scrinzi, H.~G. Muller, S.~Sukiasyan, M.~Ivanov, and O.~Smirnova,
  ``{Interpreting attoclock measurements of tunnelling times},'' {\em Nature
  Physics}, vol.~11, no.~6, pp.~503--508, 2015.

\bibitem{Shafir2012}
D.~Shafir, H.~Soifer, B.~D. Bruner, M.~Dagan, Y.~Mairesse, S.~Patchkovskii,
  M.~Y. Ivanov, O.~Smirnova, and N.~Dudovich, ``{Resolving the time when an
  electron exits a tunnelling barrier},'' {\em Nature}, vol.~485, pp.~343--346,
  2012.

\bibitem{Schultze2010}
M.~Schultze, M.~Fiess, N.~Karpowicz, J.~Gagnon, M.~Korbman, M.~Hofstetter,
  S.~Neppl, A.~L. Cavalieri, Y.~Komninos, T.~Mercouris, C.~A. Nicolaides,
  R.~Pazourek, S.~Nagele, J.~Feist, J.~Burgd{\"{o}}rfer, A.~M. Azzeer,
  R.~Ernstorfer, R.~Kienberger, U.~Kleineberg, E.~Goulielmakis, F.~Krausz, and
  V.~S. Yakovlev, ``{Delay in photoemission.},'' {\em Science (New York,
  N.Y.)}, vol.~328, no.~5986, pp.~1658--62, 2010.

\bibitem{Eckle2008a}
P.~Eckle, M.~Smolarski, P.~Schlup, J.~Biegert, A.~Staudte, M.~Sch{\"{o}}ffler,
  H.~G. Muller, R.~D{\"{o}}rner, and U.~Keller, ``{Attosecond angular
  streaking},'' {\em Nature Physics}, vol.~4, pp.~565--570, 2008.

\bibitem{Eckle2008}
P.~Eckle, A.~N. Pfeiffer, C.~Cirelli, and A.~Staudte, ``{Attosecond ionization
  and tunneling delay time measurements in helium},'' {\em Science (New York,
  N.Y.)}, vol.~322, pp.~1525--1529, 2008.

\bibitem{Goulielmakis2010}
E.~Goulielmakis, Z.-H. Loh, A.~Wirth, R.~Santra, N.~Rohringer, V.~S. Yakovlev,
  S.~Zherebtsov, T.~Pfeifer, A.~M. Azzeer, M.~F. Kling, S.~R. Leone, and
  F.~Krausz, ``{Real-time observation of valence electron motion},'' {\em
  Nature}, vol.~466, pp.~739--743, 2010.

\bibitem{klunder2011probing}
K.~Kl{\"u}nder, J.~Dahlstr{\"o}m, M.~Gisselbrecht, T.~Fordell, M.~Swoboda,
  D.~Guenot, P.~Johnsson, J.~Caillat, J.~Mauritsson, A.~Maquet, {\em et~al.},
  ``Probing single-photon ionization on the attosecond time scale,'' {\em
  Physical Review Letters}, vol.~106, no.~14, p.~143002, 2011.

\bibitem{guenot2012photoemission}
D.~Guenot, K.~Kl{\"u}nder, C.~Arnold, D.~Kroon, J.~Dahlstr{\"o}m, M.~Miranda,
  T.~Fordell, M.~Gisselbrecht, P.~Johnsson, J.~Mauritsson, {\em et~al.},
  ``Photoemission-time-delay measurements and calculations close to the 3
  s-ionization-cross-section minimum in ar,'' {\em Physical Review A}, vol.~85,
  no.~5, p.~053424, 2012.

\bibitem{Guenot2014}
D.~Guénot, D.~Kroon, E.~Balogh, E.~W. Larsen, M.~Kotur, M.~Miranda,
  T.~Fordell, P.~Johnsson, J.~Mauritsson, M.~Gisselbrecht, K.~Varjù, C.~L.
  Arnold, T.~Carette, A.~S. Kheifets, E.~Lindroth, A.~LʼHuillier, and J.~M.
  Dahlström, ``Measurements of relative photoemission time delays in noble gas
  atoms,'' {\em Journal of Physics B: Atomic, Molecular and Optical Physics},
  vol.~47, no.~24, p.~245602, 2014.

\bibitem{soifer2013spatio}
H.~Soifer, M.~Dagan, D.~Shafir, B.~D. Bruner, M.~Y. Ivanov, V.~Serbinenko,
  I.~Barth, O.~Smirnova, and N.~Dudovich, ``Spatio-spectral analysis of
  ionization times in high-harmonic generation,'' {\em Chemical Physics},
  vol.~414, pp.~176--183, 2013.

\bibitem{Pedatzur2015}
O.~Pedatzur, G.~Orenstein, V.~Serbinenko, H.~Soifer, B.~D. Bruner, a.~J. Uzan,
  D.~S. Brambila, A.~G. Harvey, L.~Torlina, F.~Morales, O.~Smirnova, and
  N.~Dudovich, ``{Attosecond tunnelling interferometry},'' {\em Nature
  Physics}, vol.~advance on, no.~August, pp.~1--6, 2015.

\bibitem{pfeiffer2012attoclock}
A.~N. Pfeiffer, C.~Cirelli, M.~Smolarski, D.~Dimitrovski, M.~Abu-Samha, L.~B.
  Madsen, and U.~Keller, ``Attoclock reveals natural coordinates of the
  laser-induced tunnelling current flow in atoms,'' {\em Nature Physics},
  vol.~8, no.~1, pp.~76--80, 2012.

\bibitem{Madsen2012}
N.~I. Shvetsov-Shilovski, D.~Dimitrovski, and L.~B. Madsen, ``Ionization in
  elliptically polarized pulses: Multielectron polarization effects and
  asymmetry of photoelectron momentum distributions,'' {\em Phys. Rev. A},
  vol.~85, p.~023428, Feb 2012.

\bibitem{dahlstrom2013theory}
J.~Dahlstr{\"o}m, D.~Gu{\'e}not, K.~Kl{\"u}nder, M.~Gisselbrecht,
  J.~Mauritsson, A.~L'Huillier, A.~Maquet, and R.~Ta{\"\i}eb, ``Theory of
  attosecond delays in laser-assisted photoionization,'' {\em Chemical
  Physics}, vol.~414, pp.~53--64, 2013.

\bibitem{Ivanov2011}
M.~Ivanov and O.~Smirnova, ``{How Accurate Is the Attosecond Streak Camera?},''
  {\em Physical Review Letters}, vol.~107, p.~213605, nov 2011.

\bibitem{Nagele2011}
S.~Nagele, R.~Pazourek, J.~Feist, K.~Doblhoff-Dier, C.~Lemell, K.~Tőkési, and
  J.~Burgdörfer, ``Time-resolved photoemission by attosecond streaking:
  extraction of time information,'' {\em Journal of Physics B: Atomic,
  Molecular and Optical Physics}, vol.~44, no.~8, p.~081001, 2011.

\bibitem{serbinenko2013multidimensional}
V.~Serbinenko and O.~Smirnova, ``Multidimensional high harmonic spectroscopy: a
  semi-classical perspective on measuring multielectron rearrangement upon
  ionization,'' {\em J. Phys. B: At. Mol. Opt. Phys.}, vol.~46, no.~17,
  p.~171001, 2013.

\bibitem{Maquet2014}
A.~Maquet, J.~Caillat, and R.~Ta{\"{\i}}eb, ``{Attosecond delays in
  photoionization: time and quantum mechanics},'' {\em Journal of Physics B:
  Atomic, Molecular and Optical Physics}, vol.~47, p.~204004, 2014.

\bibitem{Gaillac2016}
R.~Gaillac, M.~Vacher, A.~Maquet, R.~Ta\"{\i}eb, and J.~Caillat, ``Attosecond
  photoemission dynamics encoded in real-valued continuum wave functions,''
  {\em Phys. Rev. A}, vol.~93, p.~013410, Jan 2016.

\bibitem{Dahlstrom2012}
J.~M. Dahlstr{\"{o}}m, A.~L’Huillier, and A.~Maquet, ``{Introduction to
  attosecond delays in photoionization},'' {\em Journal of Physics B: Atomic,
  Molecular and Optical Physics}, vol.~45, p.~183001, sep 2012.

\bibitem{Wigner1955}
E.~Wigner, ``{Lower Limit for the Energy Derivative of the Scattering Phase
  Shift},'' {\em Physical Review}, vol.~98, pp.~145--147, apr 1955.

\bibitem{Smith1960}
F.~T. Smith, ``{Lifetime matrix in collision theory},'' {\em Physical Review},
  vol.~118, no.~1, pp.~349--356, 1960.

\bibitem{Landauer1994}
R.~Landauer and T.~Martin, ``{Barrier interaction time in tunneling},'' {\em
  Reviews of Modern Physics}, vol.~66, no.~January, pp.~217--228, 1994.

\bibitem{DeCarvalho2002}
C.~a.~a. de~Carvalho and H.~M. Nussenzveig, ``{Time delay},'' {\em Physics
  Reports}, vol.~364, pp.~83--174, 2002.

\bibitem{Yamada2004}
N.~Yamada, ``{Unified derivation of tunneling times from decoherence
  functionals},'' {\em Physical Review Letters}, vol.~93, no.~17, p.~170401,
  2004.

\bibitem{Kaushal2015}
J.~Kaushal, F.~Morales, L.~Torlina, M.~Ivanov, and O.~Smirnova, ``{Spin–orbit
  Larmor clock for ionization times in one-photon and strong-field regimes},''
  {\em Journal of Physics B: Atomic, Molecular and Optical Physics}, vol.~48,
  no.~23, p.~234002, 2015.

\bibitem{Keldysh1965}
L.~V. Keldysh, ``{Ionization in the field of a strong electromagnetic wave},''
  {\em Soviet Physics JETP}, vol.~20, no.~5, pp.~1307--1314, 1965.

\bibitem{PPT1}
A.~M. Perelomov, V.~S. Popov, and M.~V. Terent'ev, ``{Ionization of Atoms in an
  alternating electric field},'' {\em Soviet Physics JETP}, vol.~23, no.~5,
  pp.~924--934, 1966.

\bibitem{PPT2}
A.~M. Perelomov, V.~S. Popov, and M.~V. Terent'ev, ``{Ionization of atoms in an
  alternating electric field: II},'' {\em Soviet Physics JETP}, vol.~24, no.~1,
  pp.~207--217, 1967.

\bibitem{Faisal1973}
F.~H.~M. Faisal, ``{Multiple absorption of laser photons by atoms},'' {\em
  Journal of Physics B: Atomic and Molecular Physics}, vol.~6, pp.~L89--L92,
  1973.

\bibitem{Reiss1980}
H.~R. Reiss, ``{Effect of an intense electromagnetic field on a weakly bound
  system},'' {\em Physical Review A}, vol.~22, no.~5, pp.~1786--1813, 1980.

\bibitem{Torlina2012}
L.~Torlina and O.~Smirnova, ``{Time-dependent analytical R-matrix approach for
  strong-field dynamics. I. One-electron systems},'' {\em Physical Review A},
  vol.~86, p.~043408, 2012.

\bibitem{Torlina2012b}
L.~Torlina, M.~Ivanov, Z.~B. Walters, and O.~Smirnova, ``Time-dependent
  analytical r-matrix approach for strong-field dynamics. ii. many-electron
  systems,'' {\em Physical Review A}, vol.~86, no.~4, p.~043409, 2012.

\bibitem{Kaushal2013}
J.~Kaushal and O.~Smirnova, ``Nonadiabatic coulomb effects in strong-field
  ionization in circularly polarized laser fields,'' {\em Physical Review A},
  vol.~88, no.~1, p.~013421, 2013.

\bibitem{Kaushal2015b}
J.~Kaushal, F.~Morales, and O.~Smirnova, ``Opportunities for detecting ring
  currents using an attoclock setup,'' {\em Physical Review A}, vol.~92, no.~6,
  p.~063405, 2015.

\bibitem{Torlina2013}
L.~Torlina, J.~Kaushal, and O.~Smirnova, ``{Time-resolving electron-core
  dynamics during strong-field ionization in circularly polarized fields},''
  {\em Physical Review A - Atomic, Molecular, and Optical Physics}, vol.~88,
  p.~053403, 2013.

\bibitem{Torlina2014ab}
L.~Torlina, F.~Morales, H.~Muller, and O.~Smirnova, ``Ab initio verification of
  the analytical r-matrix theory for strong field ionization,'' {\em Journal of
  Physics B: Atomic, Molecular and Optical Physics}, vol.~47, no.~20,
  p.~204021, 2014.

\bibitem{Popruzhenko2008a}
S.~V. Popruzhenko and D.~Bauer, ``{Strong field approximation for systems with
  Coulomb interaction},'' {\em Journal of Modern Optics}, vol.~55, no.~16,
  pp.~2573--2589, 2008.

\bibitem{Popruzhenko2008b}
S.~V. Popruzhenko, G.~G. Paulus, and D.~Bauer, ``{Coulomb-corrected quantum
  trajectories in strong-field ionization},'' {\em Physical Review A}, vol.~77,
  p.~053409, 2008.

\bibitem{popruzhenko2014keldysh}
S.~Popruzhenko, ``Keldysh theory of strong field ionization: history,
  applications, difficulties and perspectives,'' {\em Journal of Physics B:
  Atomic, Molecular and Optical Physics}, vol.~47, no.~20, p.~204001, 2014.

\bibitem{popruzhenko2014invariant}
S.~Popruzhenko, ``Invariant form of coulomb corrections in the theory of
  nonlinear ionization of atoms by intense laser radiation,'' {\em Journal of
  Experimental and Theoretical Physics}, vol.~118, no.~4, pp.~580--586, 2014.

\bibitem{Corkum1993}
P.~B. Corkum, ``{Plasma perspective on strong field multiphoton ionization},''
  {\em Physical Review Letters}, vol.~71, no.~13, pp.~1994--1997, 1993.

\bibitem{schafer1993above}
K.~Schafer, B.~Yang, L.~DiMauro, and K.~Kulander, ``Above threshold ionization
  beyond the high harmonic cutoff,'' {\em Physical review letters}, vol.~70,
  no.~11, p.~1599, 1993.

\bibitem{multiphoton1988}
H.~van Linden van~den Heuvell and H.~Muller, {\em Multiphoton Processes
  (Cambridge Studies in Modern Optics 8), Editors: S.J. Smith and P.L. Knight}.
\newblock Cambridge University Press, 1988.

\bibitem{corkum1989above}
P.~Corkum, N.~Burnett, and F.~Brunel, ``Above-threshold ionization in the
  long-wavelength limit,'' {\em Physical review letters}, vol.~62, no.~11,
  p.~1259, 1989.

\bibitem{kuchiev1987atomic}
M.~Y. Kuchiev, ``Atomic antenna,'' {\em Soviet Journal of Experimental and
  Theoretical Physics Letters}, vol.~45, p.~404, 1987.

\bibitem{lewenstein1994theory}
M.~Lewenstein, P.~Balcou, M.~Y. Ivanov, A.~L’huillier, and P.~B. Corkum,
  ``Theory of high-harmonic generation by low-frequency laser fields,'' {\em
  Physical Review A}, vol.~49, no.~3, p.~2117, 1994.

\bibitem{salieres2001feynman}
P.~Sali{\`e}res, B.~Carr{\'e}, L.~Le~D{\'e}roff, F.~Grasbon, G.~Paulus,
  H.~Walther, R.~Kopold, W.~Becker, D.~Milo{\v{s}}evi{\'c}, A.~Sanpera, {\em
  et~al.}, ``Feynman's path-integral approach for intense-laser-atom
  interactions,'' {\em Science}, vol.~292, no.~5518, pp.~902--905, 2001.

\bibitem{smirnova2014multielectron}
O.~Smirnova and M.~Ivanov, ``Multielectron high harmonic generation: Simple man
  on a complex plane,'' {\em Attosecond and XUV Physics}, pp.~201--256, 2014.

\bibitem{smirnova2013multielectron}
O.~Smirnova and M.~Y. Ivanov, ``Multielectron high harmonic generation: simple
  man on a complex plane,'' {\em arXiv preprint: arXiv:1304.2413}, 2013.

\bibitem{Smirnova2008}
O.~Smirnova, M.~Spanner, and M.~Ivanov, ``{Analytical solutions for strong
  field-driven atomic and molecular one- and two-electron continua and
  applications to strong-field problems},'' {\em Physical Review A}, vol.~77,
  p.~033407, 2008.

\bibitem{smirnova2007coulomb}
O.~Smirnova, A.~S. Mouritzen, S.~Patchkovskii, and M.~Y. Ivanov,
  ``Coulomb--laser coupling in laser-assisted photoionization and molecular
  tomography,'' {\em Journal of Physics B: Atomic, Molecular and Optical
  Physics}, vol.~40, no.~13, p.~F197, 2007.

\bibitem{Lein2013}
J.~Zhao and M.~Lein, ``Determination of ionization and tunneling times in
  high-order harmonic generation,'' {\em Phys. Rev. Lett.}, vol.~111,
  p.~043901, Jul 2013.

\end{thebibliography}

\bibliographystyle{ieeetr}

\end{document}